\documentclass[12pt]{article}
\usepackage{graphicx,setspace,lscape,longtable}
\usepackage{epsfig,graphicx,float}
\usepackage{natbib}
\usepackage{amsmath,amsthm,amssymb,color,bm}
\usepackage[margin=1in]{geometry}
\usepackage[bottom]{footmisc}
\usepackage{indentfirst}
\usepackage{endnotes}
\usepackage{verbatim,thmtools}
\usepackage{thm-restate}
\usepackage{threeparttable}
\usepackage{comment}
\usepackage{algorithm}
\usepackage{titlesec}
\usepackage{algpseudocode}
\usepackage{multirow}
\usepackage{pdfpages}
\usepackage{dsfont}
\usepackage{subcaption}
\usepackage{xcolor}
\usepackage{rotating}
\usepackage{graphicx,wrapfig,lipsum}
\usepackage[hidelinks]{hyperref}
\usepackage{enumitem}
\usepackage{booktabs} 
\RequirePackage[mathlines]{lineno}
\usepackage{multirow}
\usepackage[titletoc]{appendix}
\bibpunct{(}{)}{;}{a}{,}{,}

\floatstyle{ruled}
\newfloat{algorithm}{tbhp}{loa}
\floatname{algorithm}{Algorithm}

\newcommand{\bxi}{\mbox{\boldmath $\xi$}}

\newcommand{\bdelta}{\mbox{\boldmath $\delta$}}

\newcommand{\tr}{\mathrm{tr}}

\newcommand{\beq}{\begin{eqnarray*}}
\newcommand{\eeq}{\end{eqnarray*}}

\def \tr {\mbox{tr}}

\def \E {\mathrm{E}}

\titleformat{\section}{\normalfont\Large\bfseries}{\thesection}{0.5em}{}
\titlespacing*{\section} {0pt}{5pt}{3pt}
\titlespacing*{\subsection} {0pt}{5pt}{2pt}
\numberwithin{equation}{section}
\theoremstyle{plain}

\newtheorem{prop}{Proposition}[section]

\theoremstyle{definition}

\newtheorem{remark}{Remark}[section]
\def\ben{\begin{equation*}}
\def\een{\end{equation*}}
\def\bea{\begin{eqnarray}}
\def\eea{\end{eqnarray}}
\def\bean{\begin{eqnarray*}}
\def\eean{\end{eqnarray*}}
\def\bep{\begin{prop}}
\def\eep{\end{prop}}
\def\bc{\begin{center}}
\def\ec{\end{center}}

\def \tr {\mbox{tr}}

\def \E {\mathrm{E}}

\def \card{\mbox{card}}

\usepackage{mathtools}

\DeclarePairedDelimiter\floor{\lfloor}{\rfloor}

\numberwithin{equation}{section}
\newtheorem{theorem}{\bf Theorem}

\newtheorem{assumption}{\bf Assumption}
\newtheorem{proposition}{\bf Proposition}

\newcommand{\blind}{1}

\begin{document}
\setstretch{1.2}
\title{\LARGE Power-Enhanced Two-Sample Mean Tests for High-Dimensional Compositional Data with Application to Microbiome Data Analysis\thanks{The preliminary result of this paper was included in the National Institutes of Health (NIH) grant proposal (1R01GM152812). Lingzhou Xue and Xiufan Yu’s research has been supported in part by the NSF grant DMS-1811552 and NIH grant R01GM152812.}}

\date{}

	\if1\blind
    {
    \author{Danning Li$^1$, Lingzhou Xue$^2$, Haoyi Yang$^2$, and Xiufan Yu$^3$ \\   $^1$Northeast Normal University, $^2$Penn State University, \\ and $^3$University of Notre Dame }
    \date{First Version: August 2023; This Version: March 2025}
    } \fi

    \if0\blind
    {
    \author{}
    } \fi

\maketitle{} 
\pagestyle{plain}

\begin{abstract}
Testing differences in mean vectors is a fundamental task in the analysis of high-dimensional microbiome compositional data. Existing methods may suffer from low power if the underlying signal pattern is in a situation that does not favor the deployed test. In this work, we develop two-sample power-enhanced mean tests for high-dimensional compositional data based on the combination of $p$-values, which integrates strengths from two popular types of tests: the maximum-type test and the quadratic-type test. We provide rigorous theoretical guarantees on the proposed tests, showing accurate Type-I error rate control and enhanced testing power. Our method boosts the testing power towards a broader alternative space, which yields robust performance across a wide range of signal pattern settings. Our methodology and theory also contribute to the literature on power enhancement and Gaussian approximation for high-dimensional hypothesis testing. We demonstrate the performance of our method on both simulated data and real-world microbiome data, showing that our proposed approach improves the testing power substantially compared to existing methods.
\end{abstract}


\setstretch{1.5}

\section{Introduction} \label{sec: introduction}
Compositional data analysis has been receiving increasing attention in several research fields such as business analytics, ecology, and microbiome over the past few years. Especially, high-dimensional compositional data are becoming increasingly available in microbiome research, ecology, and business analytics. It has been acknowledged that changes in microbial community composition are linked to key public health issues such as obesity, inflammatory bowel disease, and hypertension 
\citep{Cho12, Morgan15, Castaner18}.
Statistical inference on such data becomes vitally important for understanding the role of microbial communities in human health. Results gained from microbiome data analysis can provide insights for leading new diagnostic or therapeutic approaches in promoting human health.
One common problem of interest in analyzing high-dimensional compositional data is to test for differences in composition between different samples or experimental groups. It naturally formulates a hypothesis test on the mean of compositional data. 

Over the past decades, researchers have devoted significant efforts to the development of testing procedures for high-dimensional mean vectors \citep{bai1996effect,chen2010two,cai2014two,wang2015high,xu2016adaptive,liu2022multiple,liu2022projection, yu2022power}. 
These methods cannot be directly applied to compositional data due to the compositional nature of the data. Compositional data are characterized by the fact that the components of the data represent proportions that sum to one. The sum-to-unity constraint implies that the covariance matrix of the data is singular, which violates the eigenvalue assumptions required by most existing high-dimensional mean tests. Directly applying methods designed for unconstrained data to compositional data can lead to inaccurate or misleading conclusions \citep{aitchison1982statistical,li2015microbiome}.

A variety of mean tests on compositional data have emerged since \cite{aitchison1982statistical}, such as  \cite{srivastava2007comparison,cuesta2009projection,tsagris2017nonparametric} and others. In particular, \cite{cao2018two} extended the \textit{maximum-type test} in \cite{cai2014two} for high-dimensional compositional data. In recent years, it has been acknowledged that the testing power of many high-dimensional testing methods often depends on the sparsity level of the signal 
\citep{fan2015power,li2015joint,xu2016adaptive,he2021asymptotically,chen2022testing,yu2022power,yu2022fisher}, which is the difference between two mean vectors in the problem of mean tests. In specific, tests based on the $L_\infty$-norm of the signal (called the \textit{maximum-type tests}) tend to be powerful under sparse alternatives \citep{cai2014two,cao2018two} when only a small proportion of the covariates drive differentiation between the two mean vectors, while tests based on the (squared) $L_2$-norm of the signal (called the \textit{quadratic-type tests}) appears to exhibit strong performance under dense alternatives \citep{chen2010two,chen2019two} when the signal resides in a large number of covariates though weak within each covariate. In practical data analysis, the pattern of underlying signals is unknown in advance and misspecification can limit discovery power. It is of great importance to develop robust testing procedures that remain powerful under a variety of signal patterns.

In this work, we propose a new power-enhanced two-sample mean test for high-dimensional compositional data. 
After extending the \textit{quadratic-type test} in \cite{chen2010two} for high-dimensional compositional data, we prove that the \textit{quadratic-type test} and the \textit{maximum-type test} in \cite{cao2018two}  are asymptotically independent when the dimension can be on the nearly exponential order of the sample size. Due to this important result, we propose to use Fisher's method \citep{Fisher1925} or Cauchy combination \citep{liu2020cauchy} to relax the assumptions on the signal density and improve testing power. We further show the proposed power-enhanced tests asymptotically achieve the target size and have consistent asymptotic power under mild theoretical conditions. Moreover, we examine the finite-sample performance through numerical studies and a real-world application to examining changes in the host-microbiome community in individuals with inflammatory bowel disease.

Our theory contributes to the literature on power enhancement and Gaussian approximation for high-dimensional hypothesis testing. Specifically, we relax the Gaussian or sub-Gaussian assumption and do not need the pseudo-independence structure used in \cite{chen2010two,xu2014,li2015joint,yu2022power} and many others. We use the high-dimensional Gaussian approximation theory, instead of the martingale limit theory as in \cite{chen2010two,li2015joint,yu2022power}, to prove the asymptotic null distribution of the proposed power-enhanced tests. Also, our theory can deal with the challenges caused by the singular variance of high-dimensional compositional data. When the \textit{quadratic-type} and \textit{maximum-type tests} do not satisfy the bivariate normality assumption, we offer insights on the non-asymptotic Cauchy approximation for the tail of the null distribution (see Remark 4 in Section 3), which may have its independent significance for future research.

When we finalized this paper, we learned that \cite{jiang2024testing} independently developed an approach for testing the mean of compositional data. \cite{jiang2024testing} followed  \cite{feng2024} to consider the quadratic-type statistic in \cite{srivastava2009test} and required the dimension to be comparable with the sample size. Compared with \cite{feng2024} and \cite{jiang2024testing}, our proposed test utilizes the quadratic-type statistics introduced by Chen \& Qin (2010), allowing the data dimension to be much larger than the sample size, and we do not require the Gaussian assumption or the pseudo-independence structure, and like \cite{cao2018two}, we do not need the smallest eigenvalue to be bounded away from zero.

The rest of this paper is organized as follows. Section \ref{sec:method} introduces our proposed power-enhanced mean tests for high-dimensional compositional data. Section \ref{sec:theory} presents the asymptotic independence of maximum and quadratic-type tests and the asymptotic properties of the proposed tests. Section \ref{sec:sim} carries out simulations to validate the finite-sample properties, and Section \ref{sec:real} applies the test to real-world microbiome datasets. Section \ref{sec:disc} concludes the paper with a brief discussion. 
The proofs are presented in the supplement.

 \section{Methodology}\label{sec:method}

\subsection{Mean Tests for High-Dimensional Compositional Data}\label{subsec:prob}

Let $\bxi^{(k)} = (\xi_1^{(k)}, \ldots, \xi_{n_k}^{(k)})^\top\in \mathbb{R}^{n_k \times p}$ denote the individual compositional data matrices for each group $k \in \{1, 2\}$. By the compositional nature of the observations, each row of $\bxi^{(k)}$ sums to one. Therefore, each row of $\bxi^{(k)}$ lies in the Aitchison simplex \citep{aitchison1982statistical}:
$$\mathcal{S}^{p-1}=\{(\xi_{i1},..,\xi_{ip}):\xi_{ij} > 0 \text{ for } j \in \{1,...,p\}, \sum_{j=1}^p \xi_{ij} = 1\}.
$$ 

Assume $\xi_i^{(k)} \overset{iid}{\sim} F_{\xi}^{(k)}$, where $F_{\xi}^{(k)}$ is a distribution with mean $\mu_{\xi}^{(k)} \in \mathbb{R}^p$ and covariance $\Sigma_{\xi}^{(k)} \in \mathbb{R}^{p\times p}$. A natural question of interest is to test for differences in microbial community compositions between different groups. 
The compositional constraint imposes dependencies between multiple components of the data, which makes classical statistical tests inapplicable. To be specific, traditional testing strategies place restrictions on the covariance structure of the test samples. For example, most tests require the covariance matrix to be positive definite, which is violated in the compositional space due to the singularity in the covariance matrix of compositional data. The complex composition of the data presents a significant challenge. 

To assuage this issue, we exploit the framework of compositional data analysis to shift the assumption burden off of the observed compositional data and into a latent variable space. As a common practice in compositional data analysis, we assume the compositional variables are driven by a set of latent variables, which are known as the basis, denoted by $\eta^{(k)} = (\eta_{ij}^{(k)}) \in \mathbb{R}^{n_k \times p}$ with $\eta_{ij}^{(k)} >0$ for $k=1,2$. Given the basis, we can reconstruct the associated compositional data via normalization as follows:
 \begin{align}\label{eq:xi-eta-basis}
     \xi^{(k)}_{ij} = \frac{\eta_{ij}^{(k)}}{\sum_{j=1}^p \eta_{ij}^{(k)}} \quad \text{for}\;\; i=1, \ldots, n_k;\; j=1, \ldots, p;\; k=1,2.
 \end{align}
Let $\Delta^{(k)} = (\delta_{ij}^{(k)}) \in \mathbb{R}^{n_k \times p}$ in which $\delta_{ij}^{(k)} = \log \eta_{ij}^{(k)}$ denote the \textit{log-basis} variables. We assume $\delta_{i}^{(k)} \overset{iid}{\sim} F_{\delta}^{(k)}$, where $F_{\delta}^{(k)}$ is distribution with associated mean vector $\nu^{(k)} = \{\nu^{(k)}_1,..,\nu^{(k)}_p\}$ and covariance $\Omega^{(k)}$. A natural test in this log-basis space is to simply test the equality of $\nu^{(1)}$ and $\nu^{(2)}$; however, different basis vectors $\eta^{(k)}$ can lead to identical compositional $\xi^{(k)}$. By the compositional relationship \eqref{eq:xi-eta-basis}, given $\xi$, one is only able to recover $\eta$ up to a multiplicative constant. In \cite{aitchison1982statistical}, this indicates all the bases arising from a given compositional vector can be enumerated as $\mathcal{B}(\xi) = \{t\xi; t>0\}$. Alternatively, this yields an additive many-to-one relationship in the log-basis space $\mathcal{A}(\xi) = \{\log \xi + c \textsl{1}_p; c \in \mathbb{R}\}$. Thus, the natural testing scheme must be adapted slightly to account for this additive relationship. Two log-basis vectors $\bdelta_1$ and $\bdelta_2$ are compositionally equivalent if their components only differ by a constant $c\in \mathbb{R}$. That is, both log-bases lie within the same equivalence class $\mathcal{A}(\xi)$ \citep{cao2018two}. Let $\textsl{1}_p \in \mathbb{R}^p$ denote a vector of 1s. This yields the compositional testing framework:
\begin{align}\label{eq:hypo-equ-class}
    H_0: \nu^{(1)} = \nu^{(2)} + c\textsl{1}_p \;\;\;vs \;\;\; H_1: \nu^{(1)} \neq \nu^{(2)} + c\textsl{1}_p.
\end{align}

To construct a test on \eqref{eq:hypo-equ-class}, 
it is common to transform the compositional data via a log-ratio transformation to relax the range constraint on individual components. Methods like the additive log-ratio transformation (ALR), centered log-ratio transformation (CLR), and isometric log-ratio transformation (ILR) are often used. ALR and ILR reduce the $p$-dimensional taxon vector to $(p-1)$ dimensions, requiring a reference taxon, whereas CLR, the focus of our paper, transforms the counts by taking the log ratio of each taxon’s abundance to the geometric mean within each sample. For the observed compositional vectors $\xi_i^{(1)}$ and $\xi_i^{(2)}$, the associated CLR-transformed variables are $X_i = CLR(\xi_i^{(1)})=\log \left (\frac{\xi_{i1}^{(1)}}{g(\xi_i^{(1)})},..., \frac{\xi_{ip}^{(1)}}{g(\xi_i^{(1)})}\right)$, for $i=1, \ldots, n_1$, and  $Y_i = CLR(\xi_i^{(2)}) = \log\left (\frac{\xi_{i1}^{(2)}}{g(\xi_i^{(2)})},..., \frac{\xi_{ip}^{(2)}}{g(\xi_i^{(2)})}\right)$, for $i=1, \ldots, n_2$, 
where $g(\xi_i^{(k)}) = (\prod_{j=1}^p \xi_{ij}^{(k)})^{\frac1p}$ denotes the geometric mean of $\xi_i^{(k)}$. The centered log-ratio transformation is appealing for several reasons. The individual components of $X$ and $Y$ are no longer constrained within $(0,1]$ and are instead $X_{ij}$ and $Y_{ij}$  $ \in \mathbb{R}$ allowing us to employ traditional statistical machinery. Further, the unit constraint on $\xi$ is now a sum-to-zero constraint on the components of $X$ and $Y$. However, most useful for testing purposes is the distributional relationship between $X$ or $Y$ and the log-basis $\bdelta$. The CLR transformation is scale-invariant, allowing us to substitute $(X_i, Y_i)$  for $(\xi_i^{(1)}, \xi_i^{(2)})$  and yielding the following relationship:
\begin{align}
    X_i &= G\delta_i^{(1)}, \;  Y_i = G\delta_i^{(2)}\label{eq:linear-transformation}
\end{align}
where $G = I_p - \frac{1}{p} \textsl{1}_p \textsl{1}_p^T$ with $I_p$ denoting the $p\times p$ identity matrix and $\textsl{1}_p \in \mathbb{R}^p$ denoting a vector of only 1s. Thus the distributions of $X_i$ and $Y_i$ can be completely characterized by the distributions of $\delta_i^{(1)}$ and $\delta_i^{(2)}$ via linear transformations. Suppose $X_i\overset{iid}{\sim} H^{(1)}$ and  $Y_i\overset{iid}{\sim} H^{(2)}$, where $H^{(k)}$ characterizes a distribution with the centered log-ratio mean vector  and centered log-ratio covariance matrix such that:
\begin{align}
\mu^{(k)} = G\nu^{(k)},\; \Sigma^{(k)} = G\Omega^{(k)} G^\top.
\label{eq:linear-transformation1}
\end{align}
By construction, $G$ is a rank $p-1$ matrix with associated null space $\mathcal{N}(G) = \{x\in \mathbb{R}^p: Gx = 0\} = \{c\textsl{1}_p : c \in \mathbb{R}\}$. Using this relationship, the test on \eqref{eq:hypo-equ-class} is equivalent to a two-sample test of means as follows \citep{cao2018two}:
\begin{align}
    H_0: \mu^{(1)} = \mu^{(2)}\;\; \textrm{versus} \;\;\; H_1: \mu^{(1)} \neq \mu^{(2)}. \label{eq:hypo}
\end{align}
\cite{cao2018two} proposed the maximum compositional equivalence test assuming that $\Sigma^{(1)}=\Sigma^{(2)}$ and proved its asymptotic null distribution as a Gumbel distribution as $n_1, n_2, p\rightarrow \infty$. 

 \subsection{The Proposed Power-Enhanced Mean Tests}\label{subsec:PE-mean}
The next step is to construct a more powerful test for the problem of interest \eqref{eq:hypo} against a broader alternative space. It has been well-studied that the performance of various tests fundamentally relies on the underlying signal sparsity pattern 
\citep{fan2015power,li2015joint,xu2016adaptive,he2021asymptotically,yu2022power,yu2019innovated}. When the signal pattern in $\mu^{(1)}-\mu^{(2)}$ varies, different tests may yield distinct performance. Broadly speaking, two types of test statistics are prevalent in high-dimensional hypothesis tests: the \textit{maximum-type tests} and the \textit{quadratic-type tests}. The maximum-type mean tests construct the test statistics based on estimates of the $L_{\infty}$-norm of the difference in mean vectors, i.e., $\|\mu^{(1)} - \mu^{(2)}  \|_{\infty}$. The quadratic-type mean tests design the test statistics by utilizing estimates of the squared $L_2$-norm of the mean difference, i.e., $\|\mu^{(1)} - \mu^{(2)}  \|_{2}^2$. The maximum-type tests tend to be more powerful than quadratic-type tests under the sparse alternatives \citep{cai2014two,cao2018two} when $\mu^{(1)}-\mu^{(2)}$ have only a few non-zero components that distinguish between groups, whereas the quadratic-type tests are more powerful under the dense alternatives \citep{chen2010two,chen2019two,chen2022testing} which assume the differentiation between groups is caused by several components. When the sparsity of the alternative hypothesis is well-suited to the choice of the test statistic, there is a gain in discovery power. However, mismatches between sparsity assumptions and test statistic choice can negatively impact power substantially.

We propose combining maximum-type and quadratic-type test statistics to develop a testing framework that is more robust to improper sparsity assumptions on alternative hypotheses. Thus, our proposed strategy is more flexible in practical scenarios where there may not be strong evidence to assume a given sparsity level apriori. To begin with, we first detail the maximum-type and quadratic-type test statistics of interest. 

To encompass scenarios of sparse alternatives, we adopt the maximum-type statistic from \cite{cao2018two},
\begin{align}
    M_{n_1, n_2} &=\frac{n_1n_2}{n_1+n_2}  \underset{1\leq j \leq p}{\max}\frac{(\overline{X}_j -\overline{Y}_j )^2}
    {\hat{\gamma}_{j}}. \label{test-stat:maxtype}
\end{align}
Here $\overline{X}_j = \frac{1}{n_1} \sum_{i=1}^{n_1} X_{ij}$ and   $\overline{Y}_j = \frac{1}{n_2} \sum_{i=1}^{n_2} Y_{ij}$ are the sample means of the CLR-transformed samples $\{X_i\}_{i=1}^{n_1}$ and $\{Y_i\}_{i=1}^{n_2}$, and $\widehat{\gamma}_{j} = (n_1+n_2)^{-1} \left[\sum_{i=1}^{n_1}  (X_{ij} - \overline{X}_j )^2 + \sum_{k=1}^{n_2} (Y_{kj}-\overline{Y}_j)^2\right]$ is the pooled CLR sample variance. It can be proved that under the null hypothesis, $ M_{n_1, n_2} - 2\log p + \log\log p$ converges to a Gumbel distribution as $n_1, n_2, p\rightarrow \infty$. It is worth noting that the mathematical formula of the statistic \eqref{test-stat:maxtype} is equivalent to applying \cite{cai2014two}'s maximum test to the CLR-transformed data $\{X_i\}_{i=1}^{n_1}$ and $\{Y_i\}_{i=1}^{n_2}$. That being said, as discussed in Section 3.3. of \cite{cao2018two}, the theoretical analysis is radically different, in that the regularity conditions required by \cite{cai2014two} are not directly satisfied by the observed compositional data. 
The associated $\alpha$-level test is defined as
\begin{align}
    \Phi^M_\alpha = I(M_{n_1, n_2} \geq q^M_\alpha +2 \log p - \log \log p), \label{eq:test-Maximum}
\end{align}
where $I(\cdot)$ denotes the indicator function and $q^M_\alpha$ is the upper $\alpha$-quantile of the Gumbel distribution. The null hypothesis in \eqref{eq:hypo} is rejected by the maximum-type test when $\Phi^M_\alpha=1$.  As noted previously, $M_{n_1, n_2}$ possesses high power in the sparse alternative setting but as explored in Section \ref{sec:sim}, this gain in power rapidly disappears as the signal becomes denser.

To account for the dense setting, we consider the quadratic-type statistic $Q_{n_1, n_2}$ developed by \cite{chen2010two} to the CLR-transformed samples $\{X_i\}_{i=1}^{n_1}$ and $\{Y_i\}_{i=1}^{n_2}$. Let 
\begin{align*}
    T_{n_1, n_2} &= \frac{\sum_{i\neq j}^{n_1} X_i^TX_j}{n_1(n_1-1)}+\frac{\sum_{i\neq j}^{n_2} Y_i^TY_j}{n_2(n_2-1)}-2\frac{\sum_{i=1}^{n_1}\sum_{j=1}^{n_2} X_i^TY_j}{n_1n_2}, \notag\\
    \widehat{\sigma}_{n_1,n_2}^2 &= \frac{2}{n_1(n_1-1)}\widehat{\tr((\Sigma^{(1)})^2)}+\frac{2}{n_2(n_2-1)}\widehat{\tr((\Sigma^{(2)})^2)}+\frac{4}{n_1n_2}\widehat{\tr(\Sigma^{(1)}\Sigma^{(2)})}, \notag
\end{align*}
where the  form of $ \widehat{\sigma}_{n_1,n_2}^2$ is presented in the supplement. 
It can be verified that $T_{n_1,n_2}$ and $\widehat\sigma_{n_1,n_2}^2$  are  unbiased estimates of $\|\mu^{(1)}-\mu^{(2)}\|_2^2$ and  the variance of $T_{n_1,n_2}$ (i.e., $ \sigma_{n_1,n_2}^2 = \frac{2}{n_1(n_1-1)}  \tr((\Sigma^{(1)})^2) +\frac{2}{n_2(n_2-1)} \tr((\Sigma^{(2)})^2)+\frac{4}{n_1n_2}  \tr(\Sigma^{(1)}\Sigma^{(2)})$).
The test statistic is defined as 
\begin{align}
     Q_{n_1, n_2} = \frac{T_{n_1, n_2}}{\widehat{\sigma}_{n_1,n_2}}. \label{test-stat:quadtype}
\end{align}
The associated $\alpha$-level test is, therefore,  defined as 
\begin{align}
    \Phi^Q_\alpha = I(Q_{n_1, n_2} \geq q^Q_\alpha), \label{eq:test-Quadratic}
\end{align}
where $q^Q_\alpha$ denotes the upper $\alpha$-quantile of the standard Gaussian distribution. As such, the null hypothesis in $\eqref{eq:hypo}$ is rejected by the quadratic-type test when $\Phi^Q_\alpha = 1$. As shown in Section \ref{sec:sim}, the quadratic-type test statistic $Q_{n_2, n_2}$ achieves satisfactory power when the underlying signal density is dense and the associated performance suffers as the signal becomes sparser.

In what follows, we leverage the power of both test statistics without placing stringent assumptions on signal sparsity. To this end, we construct our power-enhanced tests by employing $p$-value combination approaches, including Fisher's method \citep{Fisher1925} and Cauchy combination \citep{liu2020cauchy}. These combination methods aggregate information from the maximum-type and quadratic-type tests to combine their respective strengths. 

Let $p_M$ be the $p$-value of $M_{n_1, n_2}$ and $p_Q$ the $p$-value of  $Q_{n_2, n_2}$.  Fisher combination test statistic, denoted by $F_{n_1,n_2}$, combines both $p$-values as 
\begin{align}
    F_{n_1, n_2} = -2(\log p_M + \log p_Q).
\end{align}
Theorem \ref{thm: asymp-indep} of Section \ref{sec:theory} proves the maximum-type statistic $M_{n_1, n_2}$ and the quadratic-type statistic $Q_{n_1,n_2}$ are asymptotically independent. Thus, under $H_0$, $F_{n_1,n_2}$ converges to the $\chi^2_4$ distribution as $n_1, n_2, p \rightarrow \infty$.
The associated $\alpha$-level test is, therefore, defined as
\begin{align}
    \Phi^F_\alpha = I(F_{n_1,n_2} \geq q^F_\alpha),
    \label{eq:test-Fisher}
\end{align}
where $q^F_\alpha$ is the upper $\alpha$-quantile of the $\chi^2_4$ distribution.  Fisher combination test rejects the null hypothesis in \eqref{eq:hypo} when $\Phi_\alpha^F = 1$.

We denote the Cauchy combination test statistic as $C_{n_1,n_2}$ and define it as follows
\begin{align}
    C_{n_1,n_2} = \omega_M \tan\{(0.5-p_M)\pi\}+\omega_Q\tan\{(0.5-p_Q)\pi\},
\end{align}
where $\omega_M$ and $\omega_Q$ are non-negative weights for the maximum-type and quadratic-type test statistics respectively, and $\omega_M+\omega_Q=1$. Under $H_0$, $p_M$ and $p_Q$ follow a $\text{Unif}(0,1)$ distribution, thus $\tan\{(0.5-p_M)\pi\}$ and $\tan\{(0.5-p_Q)\pi\}$ follow a standard Cauchy distribution. Together with the asymptotic independence shown in Theorem \ref{thm: asymp-indep},  $C_{n_1, n_2}$ converges to a standard Cauchy distribution as $n_1, n_2, p\rightarrow\infty$. Therefore, the associated $\alpha$-level test is 
\begin{align}
\Phi_\alpha^C = I(C_{n_1,n_2} \geq q_\alpha^C),\label{eq:test-Cauchy}
\end{align}
where $q_\alpha^C$ is the upper $\alpha$-quantile of a standard Cauchy distribution. The Cauchy combination test rejects the null hypothesis \eqref{eq:hypo} when $\Phi^C_\alpha=1$.

For clarity, we summarize the whole testing procedure as Algorithm \ref{algo:tests}  (in Table \ref{algo:tests}). As we will study theoretical properties in Section \ref{sec:theory} and numerical properties in Section \ref{sec:sim},  both power-enhanced tests retain the appropriate $\alpha$-level type I error rate, and achieve improved power than the maximum and quadratic tests while agnostic to the underlying signal density.

\begin{table}
    \captionsetup{justification=centering}
    \caption{Algorithm \ref{algo:tests} -- Fisher combination test and Cauchy combination test \\ for testing two-sample mean vectors of high-dimensional compositional data}
    \label{algo:tests}
    \noindent\rule{\textwidth}{0.4pt}
    \begin{algorithmic}[1] 
    \Require{individual compositional data $\xi_i^{(k)}$ for $k\in\{1,2\}$ and $i=1,\ldots, n_k$}
    \Ensure{the $\alpha$-level Fisher combination test $\Phi^F_\alpha$ and $\alpha$-level Cauchy combination test $\Phi^C_\alpha$} 
    \State Apply the centered log-ratio (CLR) transformation to compositional data $\xi_i^{(k)}$ for $k\in\{1,2\}$ and $i=1,\ldots, n_k$, and obtain $$X_i= CLR(\xi_i^{(1)}) = \log \left (\frac{\xi_{i1}^{(1)}}{g(\xi_i^{(1)})},..., \frac{\xi_{ip}^{(1)}}{g(\xi_i^{(1)})}\right)$$ and $$Y_i= CLR(\xi_i^{(2)})= \log\left (\frac{\xi_{i1}^{(2)}}{g(\xi_i^{(2)})},..., \frac{\xi_{ip}^{(2)}}{g(\xi_i^{(2)})}\right),$$ 
    where $g(\xi_i^{(k)}) = (\prod_{j=1}^p \xi_{ij}^{(k)})^{\frac1p}$ denotes the geometric mean of $\xi_i^{(k)}$.
    \State  Compute the quadratic-type statistic $Q_{n_1, n_2}$ defined by \eqref{test-stat:quadtype} and maximum-type statistic $M_{n_1,n_2}$ defined by \eqref{test-stat:maxtype}, using the CLR-transformed data $\{X_i\}_{i=1}^{n_1}$ and $\{Y_i\}_{i=1}^{n_2}$ 
    \State Calculate the $p$-value $p_Q$ of $Q_{n_1, n_2}$ and the $p$-value $p_M$ of $M_{n_1, n_2}$, i.e., 
    $$p_Q = 1-\Phi(Q_{n_1,n_2}), \quad p_M = 1-F(M_{n_1,n_2}-2\log p+\log\log p), $$
    where $\Phi(x)$ is the cdf of standard normal distribution, and $F(y)=\exp\left(-\frac{1}{\sqrt{\pi}}\exp\left(-\frac{y}{2}\right)\right)$ is the cdf of a Gumbel distribution.  
    \State  Construct the Fisher combination test statistic $$F_{n_1, n_2}=-2(\log p_Q+\log p_M)$$ and the Cauchy Combination statistic $$C_{n_1,n_2}=\frac{1}{2}[\text{tan}\{(0.5-p_Q)\pi\}+\text{tan}\{(0.5-p_M)\pi\}].$$
    \State Obtain the $\alpha$-level Fisher combination test $$ \Phi^F_\alpha = I(F_{n_1,n_2} \geq q^F_\alpha),$$
    where $q^F_\alpha$ is the upper $\alpha$-quantile of the $\chi^2_4$ distribution, and the $\alpha$-level Cauchy combination test 
    $$ \Phi^C_\alpha = I(C_{n_1,n_2} \geq q^C_\alpha),$$ 
    where $q_\alpha^C$ is the upper $\alpha$-quantile of a standard Cauchy distribution
    \end{algorithmic}
     \noindent\rule{\textwidth}{0.4pt}
\end{table}

\section{Theoretical Properties}\label{sec:theory}

We first introduce four assumptions and then present the theoretical properties. 

\begin{assumption}\label{assum: linear-in-Chen}
For $ k \in \{1,2\}$, let $\delta^{(k)}=(\delta^{(k)}_1,\ldots,\delta^{(k)}_p)'$ be a $p$-dimensional random vector with mean $\nu^{(k)}$  and covariance $\Omega $ and satisfy the following conditions:
\begin{itemize}
    \item[(i)] there is a constant $  K_4$ such that, for any $\alpha \in R^p$, 
 $$\{E|\alpha'(\delta^{(k)}-\nu^{(k)})|^4\}^{1/4}\leq K_4\{E|\alpha'(\delta^{(k)}-\nu^{(k)})|^2\}^{1/2}.
 $$
    \item[(ii)] $\E(\max_j| \delta_j^{(k)}|^3)\leq    M^3(\log p)^\frac{3}{2}$ with $(\log p)^{10}M^6=o({n_1+n_2})$.
\end{itemize}
\end{assumption}
\begin{remark}
Assumption \ref{assum: linear-in-Chen}(i) on the fourth moment was used in \cite{giessing2020bootstrapping}, and Assumption \ref{assum: linear-in-Chen}(ii) was used in \cite{CCK2016}. Assumption \ref{assum: linear-in-Chen} relaxes the Gaussian or sub-Gaussian assumption used by \cite{cao2018two,chen2019two} and \cite{feng2024}, and it also relaxes the linear multivariate model assumption widely used in the theoretical analysis of quadratic-type statistics including \cite{chen2010two,xu2014,yu2022power} and so on. Note that the pseudo-independence structure in (3.2) of \cite{chen2010two} or (4.1) of \cite{xu2014} does not allow for an elliptical model. Assumption \ref{assum: linear-in-Chen} includes the elliptical model with a finite fourth moment as a special example, where $M$ equals to a constant $C$ (or $C\sqrt{\log p}$) when $\delta_j^{(k)}$'s follow sub-Gaussian (or sub-exponential) distributions \citep{CCK2016}.
\end{remark}

\begin{assumption}\label{assum: A1A2-in-Chen2} For $k \in \{1,2\}$,  the covariance matrix of $\delta^{(k)}$ satisfy: 

\begin{itemize}
    \item[(i)] there is a constant $C$ such that
 $1/C\leq \omega_{jj} \leq C$ for $1\leq i\leq p$. 
\item[(ii)]  
 $\lambda_1( \Omega )/\sqrt{\tr(\Omega^2)} =o((\log p)^{-1- \alpha_0})$ for a constant $\alpha_0>0$ 
  and  $\lambda_{p-q}( \Omega ) >0$ with  $q=o(p)$, where $\lambda_p(\Omega )\leq \ldots\leq\lambda_1(\Omega )$ are the  eigenvalues of $\Omega^{(k)}$.
  \end{itemize}
\end{assumption}
 {

\begin{remark} 
Assumption 2(i) was used in \cite{cao2018two} to bound the variances away from zero and infinity. The condition
$ \lambda_1( \Omega )/\sqrt{\tr (\Omega^2) }\to 0 $ is sufficient for establishing the central limit theorem for quadratic-type test statistics, which was used in \cite{chen2010two}.  By requiring  Assumption 2(ii), we can derive the asymptotic independence between maximum-type and quadratic-type test statistics without Gaussian assumption. Assumption 2 relaxes those conditions in \cite{feng2024} (see their assumptions (3) and (9)) used to prove the asymptotic independence result of maximum-type test statistics and quadratic-type test statistics of \cite{srivastava2009test} under normality. Instead, we allow $\lambda_1( \Omega ) $ to be larger than $\sqrt{p} (\log p)^{-1}$, and like \cite{cao2018two}, we do not bound the smallest eigenvalue $\lambda_p( \Omega ) $ away from zero. 
\end{remark}
}

\begin{assumption}\label{assum: A1A2-in-Chen1} As $\min\{n_1,n_2\}\rightarrow\infty$, $n_1/\left(n_1+n_2\right)\rightarrow c$, for some constant $c\in (0,1)$.
\end{assumption}

This is a common assumption for the theoretical analyses of high-dimensional two-sample tests, see \cite{chen2010two,cao2018two,chen2019two,yu2022fisher,yu2022power}.

Before proceeding to the next assumption, we define some useful notations. Denote the correlation matrices of $\delta^{(k)}$ by $\mathcal{R}=(\tau_{ij} )_{p\times p}$.  For any set $\mathcal{A}$, $\card(\mathcal{A})$ denotes the cardinality of $\mathcal{A}$. For  $0<r<1$, let
$
\mathcal{V}_i(\tau,r) = \left\{1\leq j\leq p: |\tau_{ij} | \geq r \right\}
$
be the set of indices $j$ such that $\delta^{(k)}_j$   is highly correlated (whose correlation $>r$) with $\delta^{(k)}_i$  for a given $i\in\{1,\dots, p\}$. And for any $\alpha>0$, let 
$
    s_i(\tau,\alpha) = \card(\mathcal{V}_i(\tau,\left(\log p\right)^{-1- \alpha})),\ i=1,\ldots, p
$
be the number of indices $j$ in the set $\mathcal{V}_i(\tau,\left(\log p\right)^{-1- \alpha})$. Moreover, define
$
    \mathcal{W}(\tau,r) = \left\{1\leq i\leq p: \mathcal{V}_i(\tau,r) \neq \varnothing \right\}
$
such that, $\forall i\in\mathcal{W}(\tau,r)$, $\delta^{(k)}_i$ is highly correlated with some other variables of $\delta^{(k)}$.
 
\begin{assumption}\label{assum: caolinli}  The correlation matrix of $\delta^{(k)}$ satisfies the following conditions.
\begin{itemize}
    \item [(i)] There exist a constant   $\alpha_0>0$ for all $\kappa>0$, $\operatornamewithlimits{\max}\limits_{1\leq i\leq p, i\not\in \Upsilon} s_i(\tau,\alpha_0)=o(p^\kappa)$.
    \item[(ii)]
There exist a constant $0<  r_0<1$,    $\card(\mathcal{W}(\tau,r_0)) =o(p)$.
\end{itemize}
\end{assumption}
{
\begin{remark} 
\cite{cao2018two} obtained the limiting distribution of the maximum test statistic under the assumptions
$\max_{i}\sum_{j=1}^p|\tau_{ij} |^2\leq r_2$ and $\max_{i,j}|\tau_{ij} |\leq r_1<1$ (see their Conditions 2 and 3). Note that Assumption \ref{assum: caolinli} includes them as the special example when  $s_i(\tau,1)=C(\log p)^2.$
\end{remark}
}

After introducing these assumptions, we can present the main results in Theorem \ref{thm: asymp-indep}. 

\begin{theorem}\label{thm: asymp-indep}
    Given Assumptions \ref{assum: linear-in-Chen}- \ref{assum: caolinli},  under the null hypothesis $H_0$, we have
\begin{equation}\label{eq: asympindep}
P\left(Q_{n_1,n_2} \leq x,\ M_{n_1,n_2}-2\log p+\log\log p \leq y\right)\overset{d}{\longrightarrow} \Phi(x)\cdot F(y)
\end{equation}
for any $x,y\in\mathbb{R}$, as $n_1,n_2,p\rightarrow\infty$. $F(y)=\exp\left(-\frac{1}{\sqrt{\pi}}\exp\left(-\frac{y}{2}\right)\right)$ is the cdf of a Gumbel distribution, $\Phi(x)$ is the cdf of standard normal distribution.
\end{theorem}

Theorem \ref{thm: asymp-indep} shares a similar philosophy with the classical results on the asymptotic independence between the sum and the maximum of random variables. To the best of our knowledge, this is the first proof of the asymptotic independence result of the quadratic form statistic of \cite{chen2010two} and the maximum statistic of \cite{cao2018two} for testing two-sample mean vectors of high-dimensional compositional data. It is worth pointing out that our results allow the dimension to be on the nearly exponential order of the sample size and do not require the Gaussian or pseudo-independence assumption. Also, our theory can deal with the challenges caused by the singular variance of high-dimensional compositional data. 

The asymptotic independence under the null hypothesis plays an essential role in obtaining the explicit limiting null distribution of scale-invariant power-enhanced tests. Given the explicit joint distribution of   $Q_{n_1,n_2}$ and $M_{n_1,n_2}$, we proceed to present the asymptotic properties of our proposed Fisher's test and the Cauchy combination test. 
Specifically, Theorem \ref{thm: size} proved the correct asymptotic size for power-enhanced tests using Fisher or Cauchy combination.

\begin{theorem}[Asymptotic Size]\label{thm: size} 
Under the same assumptions as in Theorem \ref{thm: asymp-indep}, the Fisher's combined probability test and  the Cauchy combination test for high-dimensional compositional data achieve the accurate asymptotic size, that is, under the null hypothesis,
$$P\left(F_{n_1,n_2}>q^F_\alpha\right) \rightarrow \alpha\quad \text{and } \quad P\left(C_{n_1,n_2}>q^C_\alpha\right) \rightarrow \alpha\quad \text{as } n_1,n_2,p\rightarrow \infty.
$$
\end{theorem}

 \begin{remark}
In the literature, \cite{pillai2016unexpected} proved a surprising result that the ratio of dependent Gaussian random variables follows a standard Cauchy distribution under an arbitrary covariance matrix, and \cite{liu2020cauchy} studied the non-asymptotic approximation for the tail of the null distribution of the Cauchy combination test under the bivariate normality assumption with arbitrary correlation structures. Although $Q_{n_1,n_2}$ and $M_{n_1,n_2}$ do not satisfy the bivariate normality assumption, we want to provide some insights on the non-asymptotic Cauchy approximation for the tail of the null distribution of $C_{n_1,n_2}$ using the results in Theorems \ref{thm: asymp-indep} and \ref{thm: size}. Given the asymptotic independence result in Theorem \ref{thm: asymp-indep},  when $n_1,n_2,p$ are large enough, we have 
\begin{equation} \label{eq:probability_ratio}
\frac{P(p_Q \le \alpha, p_M \le \alpha)}{P(p_Q \le \alpha)+P(p_M \le \alpha)} \approx \frac{P(p_Q \le \alpha) P(p_M \le \alpha)}{P(p_Q \le \alpha)+P(p_M \le \alpha)} \approx \frac{\alpha}2.
\end{equation}
Thus, when $n_1,n_2,p$ are large enough and $\alpha$ tends to $0$, we have 
$$
P(p_Q \le \alpha, p_M \le \alpha) \approx o(P(p_Q \le \alpha)+P(p_M \le \alpha)),
$$
which provides a key result in the proof of Theorem 1 in \cite{liu2020cauchy} (see Step 2 in the supplemental file of \cite{liu2020cauchy}). This result indicates that the probability of $Q_{n_1,n_2}$ and $M_{n_1,n_2}$ simultaneously reaching extreme values is dominated by the probability of one of them doing so when $n_1,n_2,p$ are large enough. As a result, when $n_1,n_2,p$ are large enough and $t$ tends to $\infty$, we can follow the proof of \cite{liu2020cauchy} to show that
$$
\frac{P\left(C_{n_1,n_2}>t\right)}{P(W_0>t)} \approx 1, 
$$
where $W_0$ is a standard Cauchy random variable. This result implies that $C_{n_1,n_2}$ can still have an approximately Cauchy tail under the null hypothesis in the non-asymptotic setting. Such insights are supported by the numerical properties that we will explore in Section \ref{sec:sim}, where the Cauchy combination test more closely achieves the desired size than Fisher's method. 
 \end{remark}

Next, we will study the asymptotic power. Let $\tilde{\Omega}=(1+\frac{n_1 }{n_2})\Omega $. For any fixed $\epsilon_0>0$,
define the dense alternative $\mathcal{G}_d$ and the sparse alternative $\mathcal{G}_s$ respectively in the following: 
\begin{align}
\small
\mathcal{G}_d(\epsilon_0) & = \left\{ (\nu^{(1)}, \nu^{(2)}):  \frac{n_1^2\|G(\nu^{(1)}-\nu^{(2)})\|^4}{ n_1 (\nu^{(1)}-\nu^{(2)})' G\tilde{\Omega} G  (\nu^{(1)}-\nu^{(2)}) 
+\tr\{(G\tilde{\Omega} G)^2\}) }\geq\epsilon_0 \log n \right\}; \label{eq: G1}  \\
\mathcal{G}_s(\epsilon_0) & = \left\{ (\nu^{(1)}, \nu^{(2)}):   \max_j\frac{|(G(\nu^{(1)} - \nu^{(2)}))_j|}{\{  (G \tilde{\Omega} G )_{jj} \}^{1/2}}\geq     \sqrt{\frac{(2+\epsilon_0)\log p}{n_1} }\right\}  \label{eq: G2}
\end{align}
with $(G(\nu^{(1)} - \nu^{(2)}))_j$ and $(G\tilde{\Omega}  G)_{jj}$ denote the $j$-th element  of the vector  $ G(\nu^{(1)} - \nu^{(2)}) $ and the  $j$-th  diagonal element  of the matrix  $G\tilde{\Omega} G$.

In what follows, we provide a new Gaussian approximation result for the quadratic-type statistics $T_{n_1,n_2}$ and $Q_{n_1,n_2}$, which will be pivotal to prove the consistent asymptotic power without assuming the pseudo-independence structure of \cite{chen2010two}.

\begin{proposition}\label{thm: mean gaussian approx S}
Given Assumptions \ref{assum: linear-in-Chen}(i), 2 and 3, when $K_4$ is bounded, we have
\begin{equation}\label{eq: mGappros1}
  P\left(\frac{  T_{n_1,n_2} -\|\mu  \|^2  }{ \sigma_{n,a} }\leq x \right) \to \Phi(x)\quad \text{as } n_1,n_2,p\rightarrow \infty
\end{equation}
with $\mu=\mu^{(1)}-\mu^{(2)} $, $\sigma_{n,a}^2= \sigma_{n_1,n_2}^2+4\frac{(n_1+n_2) }{n_1n_2} \mu '\Sigma \mu$.   If $\frac{n_1^2\|\mu\|^4}{\max(n_1\mu'\Sigma\mu,\tr (\Sigma^2))} \to \infty$,
then
\begin{equation}\label{eq: amGappro1s1}
 P\left(Q_{n_1,n_2} >z_\alpha   \right)\to 1\quad \text{as } n_1,n_2,p\rightarrow \infty,
\end{equation}
with $z_\alpha $ being the upper $\alpha$-quantile of  the standard normal distribution. 
\end{proposition}

\begin{remark}
\cite{xu2014} and \cite{giessing2020bootstrapping,giessing2023bootstrap} studied Gaussian approximation for $\ell_2$-statistics in one-sample mean test for high-dimensional data. However, as pointed by \cite{chen2010two}, the $\ell_2$-statistics is in fact a biased estimator for $\|\mu^{(1)}-\mu^{(2)}\|^2$. As a result, \cite{xu2014} and \cite{giessing2020bootstrapping,giessing2023bootstrap} all required either $p/n$ to be bounded or much stronger moment conditions to handle the bias term. For example, \cite{xu2014} used the pseudo-independence structure in the linear process to justify their moment conditions, \cite{giessing2020bootstrapping} assumed a low-rank structure within their models, and \cite{giessing2023bootstrap} required $p = o(n^{\frac{1}{3}})$  with an identity covariance matrix. Our new result in Proposition \ref{thm: mean gaussian approx S} provides an enhanced flexibility and applicability in high-dimensional settings where the dimension is on the nearly exponential order of the sample size. 
\end{remark}

Now, using this new Gaussian approximation result, we show that Fisher's combined test $F_{n_1,n_2}$ can successfully boost the power against either dense or sparse alternatives. It is known that Fisher's method enjoys the Bahadur efficiency when combining independent tests. Thus, we only focus on the asymptotic power of Fisher's method in the following theorem.

\begin{theorem}[Asymptotic  Power]\label{thm: power}
Under the same assumptions as in Proposition \ref{thm: mean gaussian approx S}, Fisher's method achieves consistent asymptotic power, that is,  for any $\epsilon_0>0$, 
{
$$\inf_{(\nu^{(1)},\nu^{(2)})\in \mathcal{G}_d(\epsilon_0) \cup \mathcal{G}_s(\epsilon_0)} P\left(F_{n_1,n_2}\geq q^F_\alpha\right) \rightarrow 1 \quad \text{as } n_1,n_2,p\rightarrow \infty.$$
}

\end{theorem}

\section{Numerical Properties}\label{sec:sim}
In this section, we evaluate the empirical power and size under different testing settings. There are two types of covariance structures, under three various $(n,p)$ combinations, with two different data-generating methods. 
Aside from our proposed Fisher combination test \eqref{eq:test-Fisher} and Cauchy combination test \eqref{eq:test-Cauchy}, we compare to the Maximum test \eqref{eq:test-Maximum} and Quadratic test \eqref{eq:test-Quadratic}. To evaluate the influence of different data transformation methods, we consider an alternative max-type test by first transforming the compositional data using additive log ratio (ALR) and subsequently applying the Euclidean maximum mean test \citep{cai2014two} to the transformed data. 
Note that the two-sample test on the mean vectors of two groups can be equivalently regarded as a test of global association between the compositions and a dichotomous response. From this perspective, we include the Microbiome Regression-Based Kernel Association Test (MiRKAT) \citep{zhao2015testing} as an additional benchmark.

We consider two different data-generating frameworks: Gaussian distributions and Gamma distributions. Under the Gaussian framework, we generate the log-basis vectors from 
\begin{align*}
    \delta_i^{(1)} \sim N_p(\nu^{(1)}, \Omega) \ (\text{for } i=1,\ldots, n_1),  \quad
     \delta_i^{(2)} \sim N_p(\nu^{(2)}, \Omega) \ (\text{for } i=1,\ldots, n_2).
\end{align*}
Under the Gamma framework, we generate the log-basis vectors from 
\begin{align*}
     \delta_i^{(1)} = \nu^{(1)}+FU_i^{(1)}/\sqrt{10}  \ (\text{for } i=1,\ldots, n_1), \quad
     \delta_i^{(2)} = \nu^{(2)}+FU_i^{(2)}/\sqrt{10} \ (\text{for } i=1,\ldots, n_2),
\end{align*}
where $U_i^{(k)}$ are independent standard gamma with shape parameter ten,  $F=QS^{\frac{1}{2}}$ in which $Q$ and $S$ are determined by $\Omega = QSQ^T$ via singular value decomposition. Given the log-basis vectors $\delta_i^{(k)}$, the centered log-ratio transformed data is obtained as described in Section \ref{sec:method}. That is, $X_i = G\delta_i^{(1)}$, $Y_i = G\delta_i^{(2)}$ with $G = I_p -\frac{1}{p}\mathit{1}_p\mathit{1}_p^T$.

The above $\Omega$ are set to mimic two different dependent structures. The first is an AR(1) matrix, $\Omega^{AR} = (\Omega^{AR}_{ij})_{1\leq i,j\leq p}$ with $\Omega^{AR}_{ij} = \rho_{AR}^{|i-j|}$ and $\rho_{AR}=0.5$. The AR(1) matrix yields a dense covariance structure, though ensuring that the magnitude of the covariances rapidly decays \citep{Bickel08, Cai11}. The second is a random sparse block covariance matrix $\Omega^{BS} = (A_1,A_2)$, where $A_1 =B+\varepsilon I_q$ in which $B$ is a symmetric matrix with the lower-triangular entries drawn from $[-1,-0.5]\cup [0.5,1]$ uniformly with probability $0.5$, and $A_2=I_{p-q}$ where $q=\floor{3\sqrt{p}}$. We let $\varepsilon$ be $\max\{-\lambda_{min}(B),0\}+0.05$ to ensure the positive definiteness of the covariance matrix. The random block matrix allows for sparsity to be embedded within the covariance structure and mimics the structure in \cite*{Cao17,li2023robustT, li2023robustH}. 

We aim to design the mean vectors $(\nu^{(1)},\nu^{(2)})$ in such a way that the signals are comparable across all configurations of sample size and dimensionality. To accomplish this, we set $\nu^{(1)} = 0_p$ and choose the non-zero entries of $\nu^{(2)}$ to be equal and generated to satisfy: $||\nu^{(1)}-\nu^{(2)}||^2/\sqrt{\text{tr}(\Omega^2)} = 0.1$. To evaluate test performances in various signal sparsity settings, we consider different proportions of non-zero signals, taking values in  $\{0.01p, 0.05p, 0.2p, 0.5p\}$. The locations of non-zero entries are randomly selected.
As $\nu^{(2)}$ becomes denser, the individual signal strength of each element decreases. To evaluate the testing size, we include a case where $\nu^{(1)}=\nu^{(2)}=0_p$ for the null setting.
We set $n_1=n_2=n$ and generate the samples under three $(n,p)$-configurations: $(n,p)\in$ $\{(100,200)$, $(100,500)$, $(100,1000)\}$. We evaluate each test at the significance levels of $\alpha=0.05$ and report the empirical percentages of rejections in Table \ref{tab:simu-0.05}.


\begin{sidewaystable}
    \caption{The empirical percentages of rejection at the significance level of $\alpha=0.05$.}
    \label{tab:simu-0.05}
    \resizebox{\textwidth}{!}{
    \begin{tabular}{cccrrrrrr|rrrrrrr}
    \toprule
     \multirow{2}{*}{$\Omega$} & \multirow{2}{*}{$(n,p)$} & Percentage of 
       & \multicolumn{6}{c|}{Gaussian} & \multicolumn{6}{|c}{Gamma} \\ 
    && nonzero signals & Max &  Max$_{alr}$ & Quad & Fisher & Cauchy & MiRKAT & Max & Max$_{alr}$ & Quad & Fisher & Cauchy & MiRKAT \\
    \cmidrule(lr){1-3} \cmidrule(lr){4-9 }\cmidrule(lr){10-15}
    \multirow{15}{*}{$\Omega^{AR}$} & \multirow{5}{*}{(100, 200)} 
     & 0\% & 0.053 & 0.016 & 0.063 & 0.074 & 0.061 & 0.054 & 0.045 & 0.018 & 0.037 & 0.050 & 0.038 & 0.082 \\ 
    && 1\% & 1.000 & 0.956 & 0.934 & 1.000 & 1.000 & 0.988 & 1.000 & 0.804 & 0.339 & 1.000 & 1.000 & 0.106 \\ 
    && 5\% & 0.929 & 0.331 & 0.915 & 0.979 & 0.965 & 0.902 & 0.721 & 0.206 & 0.310 & 0.739 & 0.696 & 0.073 \\ 
    && 20\% & 0.328 & 0.135 & 0.860 & 0.865 & 0.820 & 0.752 & 0.219 & 0.078 & 0.246 & 0.358 & 0.302 & 0.065 \\  
    && 50\% & 0.159 & 0.069 & 0.524 & 0.506 & 0.467 & 0.401 & 0.129 & 0.074 & 0.178 & 0.237 & 0.208 & 0.052 \\ 
    \cline{2-15}
     & \multirow{5}{*}{(100, 500)}
     & 0\% & 0.057 & 0.031 & 0.052 & 0.083 & 0.065 & 0.049 & 0.061 & 0.021 & 0.050 & 0.060 & 0.055 & 0.059 \\ 
    && 1\% & 1.000 & 0.806 & 0.962 & 1.000 & 1.000 & 0.991 & 1.000 & 0.991 & 0.692 & 1.000 & 1.000 & 0.174 \\ 
    && 5\% & 0.784 & 0.189 & 0.943 & 0.975 & 0.955 & 0.907 & 0.989 & 0.409 & 0.628 & 0.997 & 0.993 & 0.133 \\ 
    && 20\% & 0.233 & 0.071 & 0.856 & 0.819 & 0.796 & 0.749 & 0.383 & 0.118 & 0.493 & 0.683 & 0.582 & 0.057 \\
    && 50\% & 0.122 & 0.047 & 0.525 & 0.499 & 0.447 & 0.402 & 0.165 & 0.085 & 0.254 & 0.329 & 0.276 & 0.044 \\
    \cline{2-15}
     & \multirow{5}{*}{(100, 1000)}
     & 0\% & 0.058 & 0.027 & 0.046 & 0.064 & 0.054 & 0.040 & 0.072 & 0.030 & 0.057 & 0.073 & 0.072 & 0.046 \\ 
    && 1\% & 1.000 & 0.603 & 0.950 & 1.000 & 1.000 & 0.975 & 1.000 & 0.631 & 0.957 & 1.000 & 1.000 & 0.095 \\ 
    && 5\% & 0.651 & 0.128 & 0.946 & 0.965 & 0.951 & 0.914 & 0.641 & 0.135 & 0.951 & 0.972 & 0.948 & 0.090 \\ 
    && 20\% & 0.182 & 0.062 & 0.855 & 0.820 & 0.803 & 0.751 & 0.194 & 0.062 & 0.843 & 0.817 & 0.804 & 0.082 \\ 
    && 50\% & 0.135 & 0.046 & 0.560 & 0.521 & 0.498 & 0.417 & 0.117 & 0.044 & 0.497 & 0.469 & 0.439 & 0.068 \\  
    \hline
    \multirow{15}{*}{$\Omega^{BS}$} & \multirow{5}{*}{(100, 200)} 
     & 0\% & 0.053 & 0.029 & 0.066 & 0.081 & 0.073 & 0.056 & 0.045 & 0.036 & 0.051 & 0.061 & 0.053 & 0.056 \\ 
    && 1\% & 0.996 & 0.993 & 0.946 & 0.999 & 0.999 & 0.671 & 0.998 & 0.997 & 0.949 & 0.997 & 0.997 & 0.066 \\ 
    && 5\% & 1.000 & 0.881 & 0.943 & 1.000 & 1.000 & 0.389 & 1.000 & 0.883 & 0.939 & 1.000 & 1.000 & 0.079 \\ 
    && 20\% & 0.825 & 0.385 & 0.871 & 0.979 & 0.952 & 0.246 & 0.828 & 0.351 & 0.833 & 0.967 & 0.932 & 0.049 \\  
    && 50\% & 0.314 & 0.176 & 0.524 & 0.662 & 0.575 & 0.134 & 0.304 & 0.183 & 0.498 & 0.609 & 0.513 & 0.057 \\ 
    \cline{2-15}
     & \multirow{5}{*}{(100, 500)}
     & 0\% & 0.056 & 0.042 & 0.046 & 0.069 & 0.056 & 0.037 & 0.060 & 0.038 & 0.058 & 0.067 & 0.067 & 0.040 \\ 
    && 1\% & 1.000 & 0.999 & 0.969 & 1.000 & 1.000 & 0.351 & 1.000 & 1.000 & 0.959 & 1.000 & 1.000 & 0.068 \\ 
    && 5\% & 1.000 & 0.675 & 0.951 & 1.000 & 1.000 & 0.216 & 1.000 & 0.696 & 0.933 & 1.000 & 1.000 & 0.055 \\ 
    && 20\% & 0.672 & 0.226 & 0.863 & 0.956 & 0.914 & 0.143 & 0.683 & 0.231 & 0.856 & 0.962 & 0.925 & 0.050 \\
    && 50\% & 0.261 & 0.104 & 0.521 & 0.609 & 0.513 & 0.092 & 0.275 & 0.110 & 0.518 & 0.634 & 0.532 & 0.062 \\
    \cline{2-15}
     & \multirow{5}{*}{(100, 1000)}
     & 0\% & 0.062 & 0.030 & 0.054 & 0.061 & 0.061 & 0.062 & 0.067 & 0.025 & 0.064 & 0.073 & 0.070 & 0.055 \\ 
    && 1\% & 1.000 & 1.000 & 0.960 & 1.000 & 1.000 & 0.138 & 1.000 & 1.000 & 0.963 & 1.000 & 1.000 & 0.051 \\ 
    && 5\% & 1.000 & 0.520 & 0.955 & 1.000 & 1.000 & 0.090 & 1.000 & 0.527 & 0.953 & 1.000 & 1.000 & 0.056 \\ 
    && 20\% & 0.591 & 0.171 & 0.861 & 0.949 & 0.899 & 0.076 & 0.591 & 0.181 & 0.861 & 0.949 & 0.907 & 0.055 \\ 
    && 50\% & 0.245 & 0.095 & 0.559 & 0.629 & 0.555 & 0.068 & 0.248 & 0.086 & 0.525 & 0.607 & 0.522 & 0.056 \\  
    \bottomrule
    \end{tabular}
    }

    Note: ``Max'' denotes the Maximum test \eqref{eq:test-Maximum}, which is mathematically equivalent to transforming the data using centered log-ratio (CLR) first and subsequently applying the Euclidean maximum mean test \citep{cai2014two}. 
    ``Max$_{alr}$'' denotes an alternative max-type test that transforms the data using additive log ratio (ALR) and subsequently applies the Euclidean maximum mean test. 
    ``Quad'' denotes the Quadratic test \eqref{eq:test-Quadratic}, ``Fisher'' denotes the Fisher combination test \eqref{eq:test-Fisher}, and ``Cauchy'' denotes the Cauchy combination test \eqref{eq:test-Cauchy}. 
\end{sidewaystable}

Across all covariance and data-generating setups, our proposed method is robust to the true underlying sparsity of the signal. As noted previously, the Maximum test has increased power when the signal is extremely sparse while the Quadratic test has improved power in the dense signal setting. This can be seen as the Maximum test tends to have improved power to the Quadratic test when the signal percentage 
is at the $1\%$ and $5\%$ level. However, at the $20\%$ and $50\%$ the Maximum test drastically loses power as expected. The inverse holds for the Quadratic test which has a substantial power increase at the denser signal settings but sacrifices a modicum of power in the sparse setting. However, the power loss of the quadratic test in the sparse setting is much smaller than the power loss of the maximum-type test in the dense settings. The ALR-based maximum-type test exhibits a similar pattern to that of the CLR-based maximum-type test: it tends to be powerful under sparse alternatives and has decreased power under dense alternatives, whereas ALR-based tests are empirically less powerful than CLR-based tests across all the simulation settings considered in this study. MiRKAT appears to align closely with the Quadratic test in settings with Gaussian log-basis vectors under AR covariances but shows a lack of low in the other three settings. Especially when samples are generated from Gamma log-basis vectors, MiRKAT is almost powerless.

Combination tests afford means of protection against improperly assumed sparsity. In the most sparse setting, where the Maximum test is the most effective, we note that  Fisher and Cauchy combination tests have improved power over the Quadratic test and are as powerful as the maximum-type test. While the Maximum test drastically reduces in power as the signal becomes denser, in these cases, the combination tests do not suffer due to the improved performance of the Quadratic tests. Thus, in the densest setting, we observe that the power of these methods is extremely comparable to the Maximum test and even has a higher power to the Quadratic test in sparse covariance settings. The Cauchy combination test appears to be slightly more conservative than the Fisher combination test; though, also boasts a somewhat tighter control over the type I error. The power differential between the two methods is small and the Cauchy combination method is still greatly more powerful than the Maximum test in a dense signal setting. Across all configurations, MiRKAT underperforms the proposed combination tests. It is important to note that these combination tests are specifically designed for hypothesis testing on two-sample mean vectors, whereas MiRKAT operates within the framework of a (generalized) linear regression model for the hypothesis testing on regression coefficients. Intuitively, MiRKAT is applicable to a broader range of problems that subsumes the two-sample mean test as a special case. Especially in the presence of covariates, MiRKAT is more flexible by allowing for the inclusion of covariates in the models and considering the effects of covariates when testing the regression coefficient of interest. However, such flexibility comes at a cost - when applied to the specific task of testing two-sample mean vectors, MiRKAT appears to be less powerful than other tests that are explicitly designed for this purpose. In Appendix B.1, we provide exploratory studies on covariate adjustments in two-sample mean tests. A comprehensive study and the theoretical guarantee on covariance adjustments in testing mean vectors are out of the scope of this work and shall be left for future work.

We also evaluate the impact of sample imbalance by letting $(n_1, n_2)$ take values in \{(20,180), (40,160), (60,140), (80,120), (100,100)\} and fixing $p$ at 500. The empirical percentages of rejections are summarized in Table \ref{tab:simu-sample-imbalance}. The results reveal that the size performance of the proposed tests is insensitive to class imbalance, in other words, the type-I error rate can be well controlled across all $(n_1, n_2)$ pairs. The power, however, appears to be less powerful when samples are highly imbalanced (e.g., when $(n_1, n_2)=(20,180)$) and more powerful when samples are balanced across the two groups (i.e., when $(n_1,n_2) = (100,100)$). What's more, we investigate how correlation among variables affects the test performance by considering various values of $\rho_{AR}$ in $\Omega^{AR}$. Results are presented in Appendix B.2 due to space limitations. 

\begin{table}
    \centering
    \captionsetup{justification=centering}
     \caption{The empirical percentages of rejection at the significance level of $\alpha=0.05$ \\ for the evaluation of the impact of sample imbalance.}
    \label{tab:simu-sample-imbalance}
    \resizebox{\textwidth}{!}{
    \begin{tabular}{cccrrrrrr|rrrrrrr}
    \toprule
    \multirow{2}{*}{$\Omega$} & \multirow{2}{*}{$(n_1, n_2,p)$} & Percentage of 
       & \multicolumn{6}{c|}{Gaussian} & \multicolumn{6}{|c}{Gamma} \\ 
    && nonzero signals & Max &  Max$_{alr}$ & Quad & Fisher & Cauchy & MiRKAT & Max & Max$_{alr}$ & Quad & Fisher & Cauchy & MiRKAT \\ 
    \cmidrule(lr){1-3} \cmidrule(lr){4-9 }\cmidrule(lr){10-15}
     \multirow{15}{*}{$\Omega^{AR}$} & \multirow{5}{*}{(20, 180, 200)}
       &  0\% & 0.063 & 0.021 & 0.067 & 0.082 & 0.070 & 0.044 & 0.067 & 0.022 & 0.037 & 0.063 & 0.052 & 0.053 \\
      &&  1\% & 0.778 & 0.204 & 0.316 & 0.780 & 0.767 & 0.415 & 0.780 & 0.187 & 0.249 & 0.772 & 0.742 & 0.066 \\
      &&  5\% & 0.186 & 0.060 & 0.304 & 0.365 & 0.312 & 0.292 & 0.184 & 0.038 & 0.240 & 0.321 & 0.258 & 0.090 \\
      && 20\% & 0.101 & 0.049 & 0.284 & 0.275 & 0.245 & 0.218 & 0.118 & 0.037 & 0.183 & 0.231 & 0.181 & 0.060 \\
      && 50\% & 0.079 & 0.034 & 0.158 & 0.164 & 0.147 & 0.126 & 0.085 & 0.044 & 0.112 & 0.159 & 0.114 & 0.063 \\
      \cmidrule(lr){2-15}
      & \multirow{5}{*}{(40, 160, 500)} 
  &  0\% & 0.059 & 0.019 & 0.047 & 0.063 & 0.053 & 0.050 & 0.070 & 0.025 & 0.038 & 0.073 & 0.056 & 0.051 \\
 &&  1\% & 0.992 & 0.509 & 0.692 & 0.997 & 0.993 & 0.790 & 0.996 & 0.520 & 0.654 & 0.996 & 0.994 & 0.075 \\
 &&  5\% & 0.446 & 0.118 & 0.691 & 0.773 & 0.707 & 0.641 & 0.425 & 0.107 & 0.627 & 0.728 & 0.647 & 0.098 \\
 && 20\% & 0.158 & 0.050 & 0.550 & 0.535 & 0.483 & 0.450 & 0.167 & 0.045 & 0.479 & 0.496 & 0.441 & 0.096 \\
 && 50\% & 0.097 & 0.047 & 0.279 & 0.294 & 0.245 & 0.231 & 0.100 & 0.053 & 0.262 & 0.282 & 0.249 & 0.085 \\
\cmidrule(lr){2-15}
 & \multirow{5}{*}{(60, 140, 500)} 
 &   0\% & 0.069 & 0.029 & 0.057 & 0.080 & 0.069 & 0.050 & 0.055 & 0.028 & 0.055 & 0.081 & 0.060 & 0.045 \\
 &&  1\% & 1.000 & 0.713 & 0.882 & 1.000 & 1.000 & 0.943 & 1.000 & 0.733 & 0.860 & 1.000 & 1.000 & 0.088 \\
 &&  5\% & 0.648 & 0.162 & 0.856 & 0.935 & 0.895 & 0.830 & 0.642 & 0.165 & 0.827 & 0.905 & 0.864 & 0.105 \\
 && 20\% & 0.206 & 0.066 & 0.739 & 0.736 & 0.699 & 0.633 & 0.187 & 0.056 & 0.697 & 0.678 & 0.651 & 0.099 \\
 && 50\% & 0.123 & 0.045 & 0.442 & 0.427 & 0.384 & 0.324 & 0.122 & 0.057 & 0.405 & 0.400 & 0.363 & 0.086 \\
\cmidrule(lr){2-15}
 & \multirow{5}{*}{(80, 120, 500)} 
  &  0\% & 0.066 & 0.021 & 0.049 & 0.075 & 0.061 & 0.039 & 0.057 & 0.026 & 0.046 & 0.066 & 0.047 & 0.055 \\
 &&  1\% & 1.000 & 0.798 & 0.933 & 1.000 & 1.000 & 0.976 & 1.000 & 0.826 & 0.937 & 1.000 & 1.000 & 0.078 \\
 &&  5\% & 0.755 & 0.205 & 0.925 & 0.969 & 0.942 & 0.891 & 0.754 & 0.218 & 0.916 & 0.968 & 0.950 & 0.109 \\
 && 20\% & 0.212 & 0.078 & 0.835 & 0.811 & 0.791 & 0.720 & 0.230 & 0.086 & 0.834 & 0.815 & 0.777 & 0.095 \\
 && 50\% & 0.134 & 0.042 & 0.500 & 0.499 & 0.455 & 0.379 & 0.130 & 0.059 & 0.477 & 0.465 & 0.422 & 0.082 \\
\cmidrule(lr){2-15}
 & \multirow{5}{*}{(100, 100, 500)} 
  &  0\% & 0.056 & 0.034 & 0.055 & 0.076 & 0.061 & 0.044 & 0.065 & 0.027 & 0.048 & 0.067 & 0.056 & 0.048 \\
 &&  1\% & 1.000 & 0.801 & 0.960 & 1.000 & 1.000 & 0.979 & 1.000 & 0.823 & 0.949 & 1.000 & 1.000 & 0.080 \\
 &&  5\% & 0.795 & 0.198 & 0.941 & 0.979 & 0.963 & 0.914 & 0.781 & 0.203 & 0.943 & 0.979 & 0.950 & 0.126 \\
 && 20\% & 0.229 & 0.075 & 0.833 & 0.819 & 0.792 & 0.741 & 0.228 & 0.069 & 0.856 & 0.834 & 0.819 & 0.085 \\
 && 50\% & 0.140 & 0.053 & 0.555 & 0.529 & 0.496 & 0.417 & 0.147 & 0.058 & 0.498 & 0.468 & 0.433 & 0.067 \\
\cmidrule(lr){1-15}
 \multirow{15}{*}{$\Omega^{BS}$} & \multirow{5}{*}{(20, 180, 200)} 
  &  0\% & 0.050 & 0.037 & 0.049 & 0.060 & 0.051 & 0.046 & 0.085 & 0.038 & 0.036 & 0.073 & 0.063 & 0.047 \\
 &&  1\% & 0.999 & 0.804 & 0.339 & 0.999 & 0.998 & 0.106 & 1.000 & 0.781 & 0.238 & 1.000 & 1.000 & 0.051 \\
 &&  5\% & 0.721 & 0.206 & 0.310 & 0.739 & 0.696 & 0.073 & 0.708 & 0.183 & 0.201 & 0.697 & 0.644 & 0.052 \\
 && 20\% & 0.219 & 0.078 & 0.246 & 0.358 & 0.302 & 0.065 & 0.217 & 0.086 & 0.173 & 0.298 & 0.242 & 0.059 \\
 && 50\% & 0.129 & 0.074 & 0.178 & 0.237 & 0.208 & 0.052 & 0.143 & 0.060 & 0.106 & 0.181 & 0.157 & 0.043 \\
\cmidrule(lr){2-15}
 & \multirow{5}{*}{(40, 160, 500)} 
  & 0\% & 0.066 & 0.030 & 0.058 & 0.071 & 0.068 & 0.048 & 0.056 & 0.035 & 0.043 & 0.061 & 0.050 & 0.048 \\
 && 1\% & 1.000 & 0.991 & 0.692 & 1.000 & 1.000 & 0.174 & 1.000 & 0.989 & 0.626 & 1.000 & 1.000 & 0.055 \\
 && 5\% & 0.992 & 0.409 & 0.628 & 0.997 & 0.993 & 0.133 & 0.989 & 0.400 & 0.603 & 0.988 & 0.985 & 0.063 \\
 && 20\% & 0.426 & 0.169 & 0.526 & 0.723 & 0.632 & 0.092 & 0.383 & 0.118 & 0.493 & 0.683 & 0.582 & 0.057 \\
 && 50\% & 0.187 & 0.094 & 0.295 & 0.381 & 0.324 & 0.070 & 0.165 & 0.085 & 0.254 & 0.329 & 0.276 & 0.044 \\
 \cmidrule(lr){2-15}
 & \multirow{5}{*}{(60, 140, 500)} 
  & 0\% & 0.058 & 0.019 & 0.063 & 0.070 & 0.064 & 0.053 & 0.053 & 0.033 & 0.049 & 0.066 & 0.064 & 0.045 \\
 && 1\% & 1.000 & 0.999 & 0.888 & 1.000 & 1.000 & 0.262 & 1.000 & 1.000 & 0.869 & 1.000 & 1.000 & 0.049 \\
 && 5\% & 0.999 & 0.584 & 0.865 & 1.000 & 1.000 & 0.185 & 0.999 & 0.529 & 0.866 & 1.000 & 0.999 & 0.065 \\
 && 20\% & 0.574 & 0.192 & 0.749 & 0.895 & 0.827 & 0.105 & 0.554 & 0.181 & 0.726 & 0.871 & 0.791 & 0.044 \\
 && 50\% & 0.216 & 0.109 & 0.430 & 0.513 & 0.440 & 0.090 & 0.214 & 0.102 & 0.374 & 0.450 & 0.386 & 0.050 \\
 \cmidrule(lr){2-15}
 & \multirow{5}{*}{(80, 120, 500)} 
  & 0\% & 0.050 & 0.031 & 0.054 & 0.065 & 0.062 & 0.042 & 0.036 & 0.034 & 0.052 & 0.052 & 0.050 & 0.057 \\
 && 1\% & 0.999 & 0.999 & 0.947 & 1.000 & 1.000 & 0.309 & 1.000 & 0.999 & 0.941 & 1.000 & 1.000 & 0.067 \\
 && 5\% & 1.000 & 0.636 & 0.922 & 1.000 & 1.000 & 0.205 & 1.000 & 0.677 & 0.928 & 1.000 & 1.000 & 0.062 \\
 && 20\% & 0.644 & 0.212 & 0.852 & 0.954 & 0.904 & 0.142 & 0.671 & 0.228 & 0.832 & 0.947 & 0.905 & 0.055 \\
 && 50\% & 0.268 & 0.092 & 0.502 & 0.586 & 0.501 & 0.088 & 0.237 & 0.106 & 0.476 & 0.575 & 0.483 & 0.059 \\
 \cmidrule(lr){2-15}
 & \multirow{5}{*}{(100, 100, 500)} 
  & 0\% & 0.063 & 0.023 & 0.042 & 0.064 & 0.058 & 0.044 & 0.051 & 0.040 & 0.055 & 0.067 & 0.059 & 0.051 \\
 && 1\% & 1.000 & 1.000 & 0.963 & 1.000 & 1.000 & 0.293 & 1.000 & 1.000 & 0.952 & 1.000 & 1.000 & 0.048 \\
 && 5\% & 1.000 & 0.692 & 0.967 & 1.000 & 1.000 & 0.189 & 1.000 & 0.678 & 0.944 & 1.000 & 1.000 & 0.044 \\
 && 20\% & 0.681 & 0.198 & 0.865 & 0.952 & 0.908 & 0.133 & 0.716 & 0.216 & 0.868 & 0.966 & 0.920 & 0.072 \\
 && 50\% & 0.262 & 0.141 & 0.535 & 0.627 & 0.540 & 0.095 & 0.246 & 0.125 & 0.490 & 0.586 & 0.510 & 0.051 \\
 \bottomrule
 \end{tabular}
    }
\end{table}

\section{Real Data Analysis}\label{sec:real}

We demonstrate the effectiveness of the combination testing framework through application to a gut microbiome study carried out by \cite{Morgan15}. This study investigates changes in host gene response linked to the development of inflammatory bowel disease based on alterations in an individual's gut microbiome composition. In this study, 16S rRNA is sequenced and the resulting reads are clustered at a 97\% similarity level to form the resulting operational taxonomic units (OTUs), which are proxies for the underlying microbial taxa. These OTU counts are inferred by sequencing 16S RNA and clustering the resulting reads to serve as proxies for the underlying taxa. This procedure identified over 7,000 unique bacterial taxa, though many were at extremely low abundance levels. As if common in microbiome analysis, these OTUs were then aggregated at the genera level to form $303$ genera counts. We focus on taxa of interest that had at least $10$ total counts. 
This preliminary cleaning step yields $p=226$ genera of interest. 
Microbial count data are often sparse, with some taxa absent in certain samples. To address this, we follow a common practice in the literature by adding a pseudo count of 0.5 \citep{srinivasan2021compositional}. A pseudo count is a small constant added to zero counts to facilitate statistical and mathematical computations. Specifically, zero counts are replaced with the pseudo count, transformed to relative abundances, and then subjected to the CLR transformation. Then we can construct a series of binary tests of means from the following set of associated metadata. Below, we summarize the three metadata values of interest: 
\begin{enumerate}
    \item Antibiotic (A): Did the individuals take antibiotics (YES/NO)
    \item Location (L): Location of Pouch vs Pre-pouch ileum (Pouch/PPI)
    \item Class (C): Classification of familial adenomatous polyposis(FAP) vs non-FAP (FAP/nFAP)
\end{enumerate}

Given these metadata values, we construct twelve different two-way interaction tests. Note that we omit the tests for main effects as signals for these main effects were relatively strong, and therefore, all tests indicate a relevant signal. Instead, comparing situations where the Max test and Quadratic test disagree is more attractive. Results are summarized in Table \ref{tab:testRes}.

\begin{table}[h]
\centering
\captionsetup{justification=centering}
\caption{Comparisons of testing two-way interactions of metadata using various methods. }\label{tab:testRes}
\begin{tabular}{rrrrrrrr}
\hline
\multicolumn{7}{c}{Two-way Interaction Tests}\\
  \hline
 Test Name & $n_1$ & $n_2$ & Max & Quad & Fisher & Cauchy \\ 
  \hline
 Fixing C at: nFAP, Testing: L & 122 &  28 &   0 &   0 &    0 &   0 \\ 
  Fixing C at: nFAP, Testing: A &  96 &  54 &   1 &   1 &   1 &   1 \\ 
  Fixing C at: FAP, Testing: L &  74 &  31 &   0 &   0 &    0 &   0 \\ 
  Fixing C at: FAP, Testing: A &  93 &  12 &   1 &   1 &    1 &   1 \\ 
  Fixing L at: PPI, Testing: C & 122 &  74 &   1 &   1 &   1 &   1 \\ 
  Fixing L at: PPI, Testing: A & 144 &  52 &   1 &   1 &     1 &   1 \\ 
  Fixing L at: Pouch, Testing: C &  28 &  31 &   1 &   1 &     1 &   1 \\ 
  Fixing L at: Pouch, Testing: A &  45 &  14 &   1 &   1 &    1 &   1 \\ 
  Fixing A at: NO, Testing: C &  96 &  93 &   1 &   1 &   1 &   1 \\ 
    \textbf{Fixing A at: NO, Testing: L }& 144 &  45 &   1 &     0 &   1 &   1 \\ 
  \textbf{Fixing A at: YES, Testing: C} &  54 &  12 &   0 &   1 &     1 &   1 \\ 
Fixing A at: YES, Testing: L &  52 &  14 &   0 &   0 &    0 &   0 \\ 
  \hline
\end{tabular}

{Note: The tests are constructed by fixing one metadata variable at a specific level, testing a second, while averaging over the third metadata value. A value of 0 indicates that the test fails to reject the null, while a value of 1 indicates the test rejects the null. Results are reported at the significance level $\alpha=0.01$. }
\end{table}


From the table, we observe that in all cases where either the Maximum or the Quadratic test detects a signal, both Cauchy and Fisher tests also detect the signal. There are seven cases in which all the approaches unanimously reject the null, three cases in which all the approaches unanimously fail to reject the null, and two cases in which the Maximum test and the Quadratic test disagree. 

To further evaluate the Type-I error rate control of the proposed tests, we permute the data to simulate null samples and evaluate the empirical size. We use the data for testing ``Fixing C at: nFAP, Testing: L'' (in which all the tests unanimously fail to reject the null; see Table \ref{tab:testRes}) to mimic the observations under the null, and generate permutated samples from it. We repeat the procedure 1,000 times, and report the empirical rejection rates in Table \ref{tab:real-data-permutation}. We also consider different sample size proportion settings to study the test performances concerning sample imbalance. The results show that all the tests maintain control of the type-I error rate across different sample sizes and group proportions. 

\begin{table}[h]
\centering
\captionsetup{justification=centering}
\caption{The empirical percentages of rejection of tests on permutated samples \\ at the significance level of $\alpha$ = 0.05 }\label{tab:real-data-permutation}
\begin{tabular}{rrrrr}
\toprule
$(n_1,n_2)$& Max & Quad & Fisher & Cauchy \\
\midrule
(50,100)     & 0.033   & 0.059     & 0.058  & 0.054  \\
(75,75)      & 0.017   & 0.073     & 0.066  & 0.059  \\
(100,50)     & 0.027   & 0.081     & 0.078  & 0.068  \\
\bottomrule
\end{tabular}

Note: Results are reported over 1,000 permutations. 
\end{table}

We dive deeper into the two cases when the Maximum and Quadratic tests disagree.
The first is when the antibiotic status (A) is fixed to NO antibiotic regimen and the location (L) is tested. The Maximum test identifies a signal while the Quadratic test does not. This suggests that the underlying signal may be sparse as the Maximum test tends to have higher power than the Quadratic test in a sparse setting. As noted by \cite{Morgan15}, only a few microbial clades are differentially expressed when comparing the Pouch location to PPI. While the authors note that the transcriptome may be greatly shifted between locations, the difference between the microbiome in each location is small; thus, the Maximum test is well capable of capturing this sparse signal. In this setting, both Fisher and Cauchy's methods agree with the Maximum test in capturing this sparse underlying signal.

The second is when the antibiotic status (A) is fixed to YES and the class status (C) is tested. The Maximum test is significant while the Quadratic test is not. \cite{Morgan15} notes that after accounting for antibiotic use, FAP and non-FAP classification is driven by the Escherichia, Actinobacteria, and Sutterella abundance. However, the paper also notes that generally higher levels of the Bacteroidetes phylum are linked to differences between the classes. The Bacteroidetes phylum constitutes a large portion of the gut microbiome \citep{Wexler07}, 
and as noted by \cite{Morgan15}, broad changes in this phylum are linked to developing the FAP class instead of the non-FAP class. 
Further, while antibiotics are known to decrease microbial diversity \citep{Wicher18}, members of the Bacteroidetes phylum are known to commonly carry some measure of antibiotic resistance. Therefore, it is likely many members of the Bacteroidetes phylum will remain active, continuing to drive the distinction between FAP and non-FAP individuals. This larger range of microbiota driving the FAP vs non-FAP classification suggests a dense underlying signal, thus showing higher power of the Quadratic test. By using Fisher or Cauchy combination, we retain increased discovery power of the Quadratic test in capturing the underlying dense signal. Improperly assuming the signal is sparse and applying the Maximum test would miss this outcome.

\begin{figure}
    \centering
    \begin{subfigure}{\textwidth}
        \centering
        \includegraphics[width=0.48\linewidth]{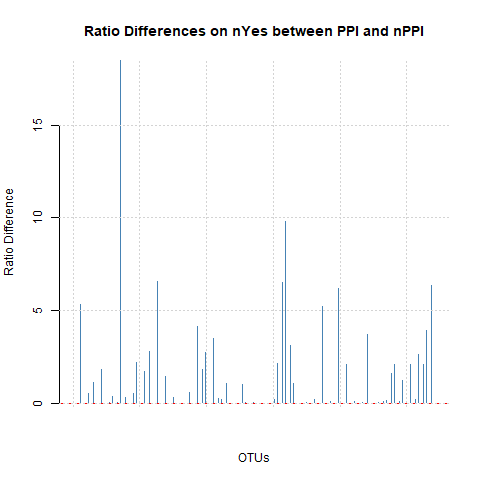}
        \includegraphics[width=0.48\linewidth]{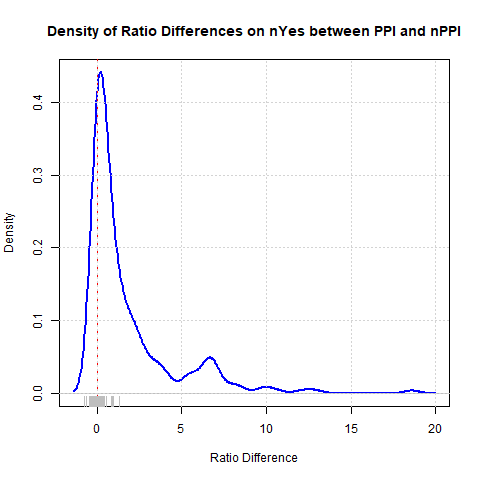}
        \captionsetup{justification=centering}
        \caption{Bar plot and density plot of the normalized mean differences \\ in the hypothesis testing  of ``Fixing A at: NO, Testing: L'' (in Table \ref{tab:testRes}) }
        \label{fig:realdata-bar-density-plots-(a)}
    \end{subfigure}

    \bigskip 
    \begin{subfigure}{\textwidth}
    \centering
    \includegraphics[width=0.48\linewidth]{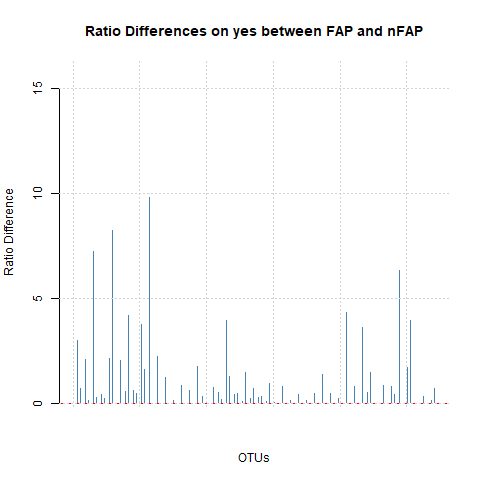}
    \includegraphics[width=0.48\linewidth]{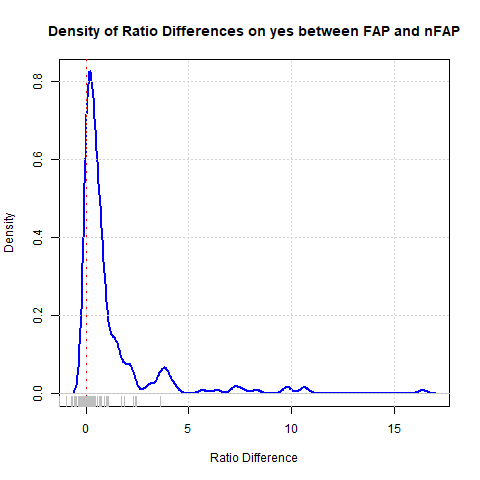}
     \captionsetup{justification=centering}
    \caption{Bar plot and density plot of the normalized mean differences \\ in the hypothesis testing of ``Fixing A at: YES, Testing: C'' (in Table \ref{tab:testRes}) }
    \label{fig:realdata-bar-density-plots-(b)}
    \end{subfigure}
    \captionsetup{justification=centering}
    \caption{Bar plots and density plots of the normalized mean differences in real data analysis}
    \label{fig:realdata-bar-density-plots}
\end{figure}

We further visualize the normalized mean differences (scaled by standard deviation) of the two cases in Figure \ref{fig:realdata-bar-density-plots}. The bar plots show the magnitude of the differences, and the density plots illustrate their empirical distributions. Figure \ref{fig:realdata-bar-density-plots-(a)} plots the case of ``Fixing A at: NO, Testing: L'' in which the Maximum test declares significance while the Quadratic test does not, while Figure \ref{fig:realdata-bar-density-plots-(b)} plots the case of ``Fixing A at: YES, Testing: C'' in which the Quadratic test declares significance while the Maximum test does not. Comparing the two cases, the largest individual standardized mean difference is notably higher in case (a), indicating that the signals are concentrated in a few OTUs with strong effects. In contrast, the mean differences in case (b) were more evenly distributed across variables, indicating weaker but denser signals across many OTUs. The signal patterns revealed by the plots are consistent with the theoretical power properties discussed in Section \ref{sec:theory}. 

\section{Conclusion}\label{sec:disc}

We proposed a power-enhanced two-sample mean test for high-dimensional compositional data.  Fisher's method \citep{Fisher1925} and Cauchy method \citep{liu2020cauchy} provide useful tools to drop restrictive signal assumptions. In doing so, we can leverage the power of maximum-type tests under sparse signals and quadratic-type tests for dense signals to improve power in compositional data analysis. Through novel theoretical derivation, we have shown the maximum-type equivalence test introduced by \cite{cao2018two} and the quadratic-type hypothesis test introduced by \cite{chen2010two} are asymptotically independent, where the dimension can be on the nearly exponential order of the sample size. We further develop a new Gaussian approximation result to prove the correct asymptotic size and consistent asymptotic power of the proposed power-enhanced tests without requiring the pseudo-independence structure. We validate numerical properties through simulation studies and a real application to study the effects of microbiome dysbiosis on host gene expression.

\newpage
\bibliographystyle{apalike}
\bibliography{paper-ref2}
\end{document}